\documentclass[english,fleqn]{article}
\usepackage{ae,aecompl}
\usepackage[T1]{fontenc}
\usepackage[latin9]{inputenc}
\usepackage[a4paper]{geometry}
\usepackage{color}
\usepackage{amsmath}
\usepackage{amssymb}
\usepackage{setspace}
\usepackage{esint}
\usepackage{float}

\makeatletter
\newcommand{\lyxaddress}[1]{
\par {\raggedright #1
\vspace{1.4em}
\noindent\par}
}

\makeatother

\usepackage{babel}
\def\be{\begin{equation}}
\def\ee{\end{equation}}
\def\bea{\begin{eqnarray}}
\def\eea{\end{eqnarray}}

\def\nn{\nonumber}

\usepackage{tikz}
\usetikzlibrary{shapes.geometric, arrows}

\tikzstyle{arriba1} = [rectangle, rounded corners, minimum width=3cm, minimum height=2cm,text centered, text width=3cm, draw=black]%, fill=red!50]

\tikzstyle{arriba2} = [circle, minimum width=3cm, minimum height=2cm, text centered, text width=3cm, draw=black]%, fill=blue!50]

\tikzstyle{abajo1} = [ellipse, minimum width=6cm, minimum height=2cm, text centered, text width=3cm, draw=black]%, fill=red!30]

\tikzstyle{abajo2} = [ellipse, minimum width=6cm, minimum height=2cm, text centered, text width=3cm, draw=black]%, fill=blue!40]

\tikzstyle{flecha} = [ultra thick,->,>=stealth]

\tikzstyle{ident} = [ultra thick,<->,>=stealth]
\begin{document}

\title{\textbf{$SU(2)$ particle sigma model: the role of non-point symmetries in global quantization}}

%\title{\textbf{Quantization of a particle on the sphere $S^{3}$: a non-canonical approach}}

\author{ V. Aldaya$^{a}$, J. Guerrero$^{a,\, b}$, F.
F. L\'opez-Ruiz$^c$ and  F. Coss\'\i o$^{a}$}

\maketitle
\begin{singlespace}

\lyxaddress{\noindent \begin{center}
\emph{$^{a}$Instituto de Astrof\'\i sica de Andaluc\'\i a (IAA-CSIC), Glorieta
de la Astronom\'\i a, E-18080 Granada, Spain}\\
\emph{$^{b}$Departamento de Matem\'atica Aplicada, Universidad de Murcia,
Campus de Espinardo, E-30100 Murcia, Spain}\\
\emph{$^{c}$Departamento de F\'\i sica Aplicada, Universidad de C\'adiz,
Campus de Puerto Real, E-11510 Puerto Real, C\'adiz, Spain}\\
\emph{}\\
\emph{valdaya@iaa.es~~ juguerre@um.es~~ paco.lopezruiz@uca.es~~ fcossiop@gmail.com}
\par\end{center}}
\end{singlespace}
\begin{abstract}
In this paper we achieve the quantization of a particle moving on
the $SU(2)$ group manifold, that is, the three-dimensional sphere
$S^{3}$, by using group-theoretical methods. For this purpose, a
fundamental role is played by contact, non-point symmetries, i.e.,
symmetries that leave the Poincar\'e-Cartan form semi-invariant at
the classical level, although not necessarily the Lagrangian.
Special attention is paid to the role played by the basic quantum
commutators, which depart from the canonical, Heisenberg-Weyl
ones, as well as the relationship between the integration measure
in the Hilbert space of the system and the non-trivial topology of
the configuration space. Also, the quantization on momentum space
is briefly outlined.
\end{abstract}

\section{Introduction}

% One of the main problems of theoretical physics today is to describe
% systematically the quantum mechanics of nonlinear dynamical systems
% with non trivial topology. This is still an open issue under discussion.
% Various attempts for solution have been made, but none of them can
% be considered as definitive to date. An example of this situation
% is the problem of the quantization of the gravitational field. Gravitation
% from the classical viewpoint is a highly nonlinear system and we would
% like to note that is this nonlinearity what is the basis of the so-called
% quantum gravity problem. I.e., is not a problem of gravitation itself
% but of quantum mechanics in general.

One of the main problems which faces theoretical physics today is
the proper quantization of  non-linear dynamical systems evolving
on a configuration space with non-trivial topology. In fact, the
difficulties encountered in  performing the quantization of
paradigmatic non-linear systems, like gravitational fields, is
traditionally associated with inherent incompatibilities between
Quantum Mechanics and the corresponding interaction. The only true
assertion in this respect seems to be that Canonical Quantization
is incompatible with non-linearity in general. But this is a very
well-known fact established at  the earliest stage of Quantum
Mechanics through the No-Go theorems \cite{No-Go} (see
\cite{Guillemin-Sternberg} for a review).

Since the very beginning, symmetry principles have constituted a
helpful tool in describing quantum phenomena
\cite{Weyl-Wigner-Waerden}, although they were focussed mostly on
the reconstruction of new solutions related to a given one by
symmetry transformations. A deeper use of symmetries were made in
more recent approaches to quantization, like Geometric
Quantization \cite{Souriau,Kostant}, based on an extension to
Physics of the geometric technique of group representation on
co-adjoint orbits of Lie Groups \cite{Kirillov}. However, those
geometric methods found serious limitations concerning the type of
operators which were compatible with the process of reduction of
the quantum representation, that is, the ``polarization'' of the
wave function.

% Since the very beginning of quantum mechanics symmetry principles
% have been an important tool in describing quantum phenomena \cite{Weyl Wigner Waerden}.
% An example of special relevance is its application in elementary particle
% physics where their contributions could hardly be underestimated.
% It is well known that canonical quantization finds serious difficulties
% in describing systems beyond the linear or quadratic \cite{Guillemin Sternberg}.
% Likewise, the success of the geometric quantization of co-adjoint
% orbits of a Lie group \cite{Co-adjoint orbits} is not guaranteed
% since when imposing the polarization conditions not all operators
% can be quantized. Therefore it is imperative to the availability of
% a method that could go beyond the above. Our approach for quantization
% is based on a previously introduced Group Approach to Quantization
% (see \cite{Aldaya review GAQ} and references there in) that generalizes
% and improves, in several aspects, the co-adjoint orbit method. Taking
% as starting point the group of symmetries of the system and using
% geometric and algebraic properties of the group this method constructs
% consistent and unambiguously an irreducible and unitary representation
% of the group. This method has proven its power and advantages several
% times. An example of this is the quantization of all orbits of the
% Virasoro group \cite{Aldaya Virasoro}.

In this paper we face the quantum description of an intrinsically
non-linear system with configuration space bearing a non-trivial
topology and, nevertheless, simple enough so as to be solved
exactly. This is a highly symmetric system, that of a free
particle evolving on the sphere $S^3$, which will allow us to
employ a Group Approach to Quantization (see
\cite{Aldaya-review-GAQ} and references therein; see especially the
pioneer papers on this subject \cite{23,Ramirez}) improving the mentioned
Kostant-Kirillov-Souriau geometric method. Taking as starting
point the group of generalized symmetries (in the sense that it
includes non-point symmetries of the Poincar\'e-Cartan form) of
our system and, using intrinsic geometric and algebraic structures
of Lie groups, we shall construct consistently and unambiguously
the unitary and irreducible representations of the basic symmetry
group, which will constitute eventually the possible quantum
representations of the physical system. This method has proven
very efficient in several non-trivial (even infinite-dimensional)
systems; a significant example was the quantization of all orbits
of the Virasoro group \cite{Aldaya-Virasoro}, appearing in 2D
quantum Gravity. The interest of the present example goes beyond
the mere study  of the quantum dynamics of a particle on a
Riemannian manifold; it will constitute a toy model for the highly
non-trivial task of properly quantizing Non-Linear Sigma Models in
field theories and, in particular, the boson sector of massive
Yang-Mills theories. In fact, the (Lie algebra of the)  local
version of the relevant symmetry here encountered, to be named,
local Sigma SU(2) group, was already pointed out in
\cite{Conferencia-Granada}.

The paper is organized as follows. In Sec. 2 we describe the basic
symmetries of the classical system of a particle moving on a
Riemannian manifold paying special attention to the non-point
character of part of them, precisely those generalizing the boosts
in the case of the flat geometry. The particular case of the free
particle moving on $S^3$ is considered in Subsec. 2.2. In Sec. 3
we perform the quantization of the
%solution manifold of the
Particle-Non-Linear-Sigma-Model associated with $SU(2)$, that is,
the free particle moving on the $SU(2)$ group manifold as
configuration space, according to a non-canonical Group Approach
to Quantization, briefly introduced in Subsec 3.1. The complete
quantization is performed in configuration space, and the
eigenvectors of the Hamiltonian operator, which turns out to be
the Laplace-Beltrami operator for $S^3$, are computed. The
realization in momentum space is also briefly discussed. Sec 4 is
devoted to remarking different non-trivial aspects which appear as
related to both non-trivial topology and non-canonical basic
commutation relations.

\section{Classical description and Symmetries}

We are interested in fundamental (elementary) systems defined by
an action. The relevant symmetries for classical mechanics are the
Noether-contact symmetries, i.e., those which leave semi-invariant
the Poincar\'e-Cartan form of the system \cite{Aldaya-Nuovo}. So
often it is not noticed that in the proof of her theorem, Noether
considered symmetries beyond point symmetries \cite{Olver1993}.
One of our objectives in this article is to show the importance
and necessity of incorporating, naturally, non-point symmetries at
the classical and quantum level.

The particle moving on the sphere $S^{3}$ is an example of what is
known in physics by a particle sigma model. Our interest in this
model lies in two aspects that make it particularly relevant.
Firstly, it is a typical non-linear model enclosing the main
problems that may be relevant to quantum mechanics and, secondly,
the topology of the configuration space is non-trivial.

\subsection{Lagrangian formalism on Riemannian manifolds}

The general particle nonlinear sigma model (PNL$\sigma$m) over a
Riemann manifold (the target manifold) is defined, traditionally,
by the following action,

\begin{equation}
S=\int dt L=\int dt\,\frac{1}{2}g_{ij}(x)\,\dot{x}^i\dot{x}^j,\,\label{action}
\end{equation}
where $L$ is the Lagrangian, $\dot{x}^i=\frac{dx^i}{dt}$, $t$ is the parameter
of the curve, $g_{ij}(x)$ is the metric of the manifold and,
as usual, the connection on the manifold is the Levi-Civita connection
(Christoffel symbols). From this action, and according to the Ordinary Hamilton Principle (OHP), we derive the
Euler-Lagrange equations of motion:
\begin{equation}
\dot{x}^i\nabla_i\dot{x}^j\equiv\ddot{x}^j+\Gamma_{kl}^j\dot{x}^k\dot{x}^l
=0,
\end{equation}
where $\nabla_i$ is the covariant derivative. These equations
are the geodesics equations of the manifold. The canonical momentum
and the Hamiltonian function are
\begin{equation}
\frac{\partial L}{\partial\dot{x}^i}=p_i=g_{ij}\dot{x}^j,
\end{equation}
and
\begin{equation}
H=\frac{\partial L}{\partial\dot{x}^i}\dot{x}^i-L=\frac{1}{2}g^{ij}p_ip_j=\frac{1}{2}g_{ij}\dot{x}^i\dot{x}^j=\frac{1}{2}\mid\dot{x}\mid^{2}.
\end{equation}
Obviously $H$ is a constant of motion.

% The Poincaré-Cartan form \cite{Aldaya-Nuovo}
% for this system is,
% %
% \begin{equation}
% \Theta_{PC}=\frac{\partial L}{\partial\dot{x}^i}(dx^i-\dot{x}^idt)+Ldt=g_{ij}\dot{x}^jdx^i-\frac{1}{2}g_{ij}\dot{x}^i\dot{x}^jd\tau=p_idx^i-Hdt.
% \end{equation}

Also in the framework of the OHP, the symmetries of any system
defined by a Lagrangian are obtained by imposing the condition of
invariance of the action, which in this case corresponds with the
semi-invariance of the action integrand, that is, invariance up to
a total derivative. If $X$ is the generator of a one-parameter
group of transformations of the space $(t, x^i, \dot{x}^j)$
\be
 X=X^t\frac{\partial}{\partial t}+X^i\frac{\partial}{\partial x^i}+\dot{X}^i\frac{\partial}{\partial\dot{x}^i}.\label{symmetry}
\ee
then this condition
is
\begin{equation}
L_XL+L\frac{dX}{dt}=\frac{df_X}{dt},
\end{equation}
where $L_X$ indicates the Lie derivative with respect to the
vector field $X$, the operator $\frac{d}{dt}$ is a total
derivative and $f_X$ is some function which depends on $X$.

The explicit form of the generator depends on the type of symmetry
that we are looking for. Usually, and again in the framework of
the ordinary variational calculus, the components $X^t(t),
X^i(x,t)$ correspond to the infinitesimal version of a
transformation on $x^i$ and perhaps the evolution parameter $t$:
\[ t'=t + \tau(t)\;,\;\;x'^i= x^i+\xi^i(x,t)\]
whereas the component $\dot{X}^i$ is the infinitesimal action on $\dot{x}^i$ induced from those of $x^i$ and $t$, that is:
\be
 \dot{X}^i=\frac{\partial X^i}{\partial t}+\frac{\partial X^i}{\partial x^k}\dot{x}^k-\frac{\partial X^t}{\partial t}\dot{x}^i\label{jet}
\ee

A particularly interesting case is that of geometric transformations where the parameter $t$ is not transformed and $\xi$ does not
depend on $t$ so that $\dot{X}^i$ is reduced to
\[\dot{X}^i=\frac{\partial X^i}{\partial x^k}\dot{x}^k \,.\]
Now the Lie derivative of $L$ with respect to $X$ results in
\be
L_XL=X^k\frac{\partial g_{ij}}{\partial x^k}+\frac{\partial X^k}{\partial x^i}g_{kj}+
\frac{\partial X^k}{\partial x^j}g_{ik} =\frac{1}{2}(L_{X}g_{ij})\,\dot{x}^i\dot{x}^j\,,
\ee
where by $L_{X}g_{ij}$ we mean the traditional expression in Riemannian geometry
\begin{equation}
L_{X}g_{ij}=X^k\frac{\partial g_{ij}}{\partial x^k}+\frac{\partial X^k}{\partial x^i}g_{kj}+\frac{\partial X^k}{\partial x^j}g_{ik}
\end{equation}
The invariance condition thus reads
\begin{equation}
L_{X}g_{ij}=0=\nabla_iX_{j}+\nabla_jX_i.
\end{equation}

These equations are known as the Killing equations and the vectors
$X^i(x)$ as the Killing vectors. This shows that the isometries of the metric
are particular symmetries of the associated  Lagrangian system. Therefore, it is then manifest that all
sigma models have, at least, these t-independent point (geometric) symmetries.

Evidently, isometries are not the only point symmetries we could
have considered. Symmetries (\ref{symmetry}) with the
infinitesimal action on $\dot{x}^i$ given by (\ref{jet}) are
referred to as point symmetries and we say that (\ref{symmetry})
is the jet prolongation of the generator of the action on just $t,
x^i$, $X^t\frac{\partial}{\partial t}+ X^i\frac{\partial}{\partial
x^i}$ \cite{Aldaya-Nuovo,Olver1993,Ibragimov1985}. Killing
symmetries are then point symmetries independent of $t$ and
preserving $t$.

Point symmetries do not exhaust, however, all symmetries required
to parameterize the set of solutions of the system described by
(\ref{action}), by means of the corresponding Noether invariants. More general, contact, non-point symmetries are
required (save for the linear case corresponding to the trivial
metric). They are defined, in the framework of the Modified
Hamilton Principle (MHP) \cite{Goldstein,Aldaya-Nuovo} as those
generated by vector fields with the general form (\ref{symmetry}),
although with general $\dot{X}^i$ (not satisfying (\ref{jet}))
leaving (semi-)invariant the Poincar\'e-Cartan form
\begin{equation}
\Theta_{PC}=\frac{\partial L}{\partial\dot{x}^i}(dx^i-\dot{x}^idt)+Ldt=g_{ij}\dot{x}^jdx^i-\frac{1}{2}g_{ij}\dot{x}^i\dot{x}^jd\tau=p_idx^i-Hdt.
\end{equation}

Contact symmetries provide those symmetries generalizing the
``boost'' of the trivial-metric system, that is, free particle in
Euclidean space (with Noether invariant the initial ``position'').
We could say that Killing symmetries only provide the generalized
momenta but generalized position are, in general, of non-point
character and have been traditionally  hidden somehow.

% The most general form of the infinitesimal generator of
% a point symmetry is \cite{Olver1993,Ibragimov1985}
%
% \begin{equation}
% X=\xi(\tau,x^{\mu})\frac{\partial}{\partial\tau}+\xi^{\alpha}(\tau,x^{\mu})\frac{\partial}{\partial x^{\alpha}}+jet\, prolongations.
% \end{equation}
% Even more important for our purposes are contact symmetries whose
% infinitesimal generators are of the form
% \begin{equation}
% X=\xi(\tau,x^{\mu},\dot{x}^{\mu})\frac{\partial}{\partial\tau}+\xi^{\alpha}(\tau,x^{\mu},\dot{x}^{\mu})\frac{\partial}{\partial x^{\alpha}}+\eta^{\alpha}(\tau,x^{\mu},\dot{x}^{\mu})\frac{\partial}{\partial\dot{x}^{\alpha}}+jet\, prolongations.
% \end{equation}
% The condition of semi-invariance of the action now takes the form
% \begin{equation}
% L_{X}\Theta_{PC}=di_{X}\Theta_{PC}+i_{X}d\Theta_{PC}=d\Lambda,
% \end{equation}
% where $d$ is exterior derivative, $i_{X}$ the interior product and
% $\Lambda$ some function. Contact symmetries play a key role in this
% article.

\subsection{Particle on the $SU(2)$ group manifold: $S^3$}
\label{particleS3}

One example containing all the ingredients of PNL$\sigma$M (in spaces
of constant curvature) is that of a particle moving on a Lie group
manifold $G$. In general we would be interested in semi-simple groups. For these
groups we can define a two-side (chiral) invariant metric, i.e., invariant by the left and
right action of the group, as follows \cite{Ketov,Isham1984}:
\begin{equation}
g_{ij}=k_{ab}\theta_{i}^{(a)}\theta_{j}^{(b)};\,\, a,b,i,j=1,\,...,dim(G),
\end{equation}
where $k_{ab}=C_{a\, j}^{\, i}C_{b\, i}^{\, j}$ is the Cartan-Killing
metric (for semi-simple algebras this metric is nonsingular), $C_{k\, j}^{\, i}$
are the structure constants of the Lie algebra of the group, and
$\theta_{i}^{(a)}$ the left/right invariant 1-forms of Cartan\footnote{Notice that the canonical 1-forms are a particular case of vierbeins $e^{(a)}_i$ as defined in \cite{vierbein}}.
The Lagrangian driving the motion on $G$ is given by
\begin{equation}
L=\frac{1}{2} m\,k_{ab}\theta_{i}^{L\,(a)}\theta_{j}^{L\,(b)}\dot{g}^{i}\dot{g}^{j}\;=\;\{L\rightarrow R\},\label{Lagrangian}
\end{equation}
where $g^{i}$ are local coordinates of the group and $m$ is the mass of the particle. As a general fact, the equations of motion can be written as
\[ \frac{d}{dt}\theta^{L\,(a)}_i\dot{g}^i\equiv\frac{d}{dt}\theta^{L\,a}= 0\,.\]

Here, we are particularly interested in the case in which $G$ is
the group $SU(2)$, the universal covering of the ordinary rotation group $SO(3)$. Thus,  the configuration
space of the particle is $S^{3}$ and the phase space is its
tangent space $T(S^{3})$. We shall parameterize locally the group
manifold with coordinates $\left\{ \epsilon^{i};\, i=1,2,3\right\}
$, where $\vec{\epsilon}$ determines the axis of rotation and
$\mid\vec{\epsilon}\mid=R\,\sin\frac{\varphi}{2}$, $\varphi$ being
the corresponding rotation angle; we make explicit the radius $R$
of the sphere where the particle evolves. In fact, when completed
with $R \rho(\vec{x})\equiv R \sqrt{1-\frac{\vec{x\,}^{2}}{R^2}}$,
these coordinates are the restriction of Cartesian coordinates on
a $\mathbb R^4$ Euclidean space to the three-dimensional sphere of
radius $R$.
 With this parametrization
%Then the group law for this group is
%
% \begin{equation}
% \epsilon^{\prime\prime i}=\sqrt{1-\frac{\vec{\epsilon\,}^{2}}{4}}\epsilon^{\prime i}+
% \sqrt{1-\frac{\vec{\epsilon\,}^{\prime2}}{4}}\epsilon^{i}+\frac{1}{2}\eta_{\cdot jk}^{i}\epsilon^{\prime j}\epsilon^{k},
% \end{equation}
%
we can derive the right- and left-invariant canonical 1-forms,
\begin{equation}
\theta^{R\,(i)}=\theta_{j}^{R\,(i)}d\epsilon^{j}=\left(\rho(\vec{\epsilon}\,)\delta_{j}^{i}+
\frac{\epsilon^{i}\epsilon_{j}}{R^2\rho(\vec{\epsilon}\,)}+\frac{1}{R}\eta_{\cdot jk}^{i}\epsilon^{k}\right)\, d\epsilon^{j},
\end{equation}
\begin{equation}
\theta^{L\,(i)}=\theta_{j}^{L\,(i)}d\epsilon^{j}=\left(\rho(\vec{\epsilon}\,)\delta_{j}^{i}+
\frac{\epsilon^{i}\epsilon_{j}}{R^2\rho(\vec{\epsilon}\,)}-\frac{1}{R}\eta_{\cdot jk}^{i}\epsilon^{k}\right)\, d\epsilon^{j},
\end{equation}
where the function $\rho$ is defined above, $\eta_{.jk}^i$ is the Levi-Civita
symbol in $3$-dimensions, and the Cartan-Killing metric, or chiral metric, on the group
\begin{equation}
k_{ij}=\frac{2}{R}\eta_{\cdot im}^{l}\frac{2}{R}\eta_{\cdot jl}^{m}=-\frac{8}{R^2}\delta_{ij},
\end{equation}
\begin{equation}
-\frac{R^2}{8}k_{kl}\theta_{i}^{R\,(k)}\theta_{j}^{R\,(l)}=g_{ij}=\left(\delta_{ij}+\frac{\epsilon_{i}\epsilon_{j}}{R^2\rho(\vec{\epsilon}\,)^2}\right).\label{metrica}
\end{equation}
The dual vectors
\begin{equation}
Z_{(i)}^{R}=Z_{(i)}^{R\, k}\frac{\partial}{\partial\epsilon^{k}}=\left(\rho(\vec{\epsilon}\,)\delta_{i}^{k}+\frac{1}{R}\eta_{\cdot ij}^{k}\epsilon^{j}\right)\,\frac{\partial}{\partial\epsilon^{k}},\label{simetriaR}
\end{equation}
\begin{equation}
Z_{(i)}^{L}=Z_{(i)}^{L\, k}\frac{\partial}{\partial\epsilon^{k}}=\left(\rho(\vec{\epsilon}\,)\delta_{i}^{k}-\frac{1}{R}\eta_{\cdot ij}^{k}\epsilon^{j}\right)\,\frac{\partial}{\partial\epsilon^{k}},\label{simetriaL}
\end{equation}
constitute the infinitesimal generators of the left and right action of the group on itself, respectively, and will turn out to be the
Killing vectors for the Cartan-Killing metric (see below). The Lie algebra (of the right-invariant vector fields) for $SU(2)$ is
\begin{equation}
\left[Z_{(i)}^{R},Z_{(j)}^{R}\right]=-\frac{2}{R}\eta_{\cdot ij}^{k}\, Z_{(k)}^{R}.
\end{equation}

From the formulas above, we write the Lagrangian for the $S^{3}$
PNL$\sigma$M:
%(we write it in terms of right objects for the sake of concreteness),
%
\begin{equation}
L=%\frac{1}{2}m\frac{R^2}{8}k_{kl}\theta_{i}^{R\,(k)}\theta_{j}^{R\,(l)}\dot{\epsilon}^{i}\dot{\epsilon}^{j}=
\frac{1}{2}m\, g_{ij} \dot{\epsilon}^{i}\dot{\epsilon}^{j}
=\frac{1}{2}m\,\left(\delta_{ij}+
\frac{\epsilon_{i}\epsilon_{j}}{R^2\rho(\vec{\epsilon\,})^2}\right)\dot{\epsilon}^{i}\dot{\epsilon}^{j}\,\,\,\,\,\,(\dot{\epsilon}^{i}=
\frac{d\epsilon^{i}}{dt}).
\end{equation}
Note that this expression can also be obtained by constraining a 4D Lagrangian to the sphere $S^3$.
%This shows that the particle is moving in space of constant curvature
%with curvature $k=\frac{1}{4}$ , i.e., the sphere of radius $R=2$
%and, therefore, in a non trivial topology.
The Poincar\'e-Cartan form,
the momentum and the Hamiltonian are \cite{Aldaya-Nuovo}
\begin{equation}
\Theta_{PC}=\frac{\partial L}{\partial\dot{\epsilon}^{i}}(d\epsilon^{i}-\dot{\epsilon}^{i}dt)+L\, dt=p_{i}d\epsilon^{i}-H\, dt,\label{thetaPC}
\end{equation}
\begin{equation}
p_{i}=\frac{\partial L}{\partial\dot{\epsilon}^{i}}=mg_{ij}\dot{\epsilon}^{j}=m\theta^{R(k)}_i\theta^R_k,
\end{equation}
\begin{equation}
H=\frac{\partial L}{\partial\dot{\epsilon}^{i}}\dot{\epsilon}^{i}-L=\frac{1}{2}m\,g_{ij}\dot{\epsilon}^{i}\dot{\epsilon}^{j}=
\frac{1}{2m}g^{-1ij}p_{i}p_{j}=\frac{1}{2}m\,\delta_{ij}\theta^{Ri}\theta^{Rj},\label{Hamiltoniana}
\end{equation}
where $g^{-1ij}=\delta^{ij}-\frac{1}{R^2}\epsilon^{i}\epsilon^{j}$ is
the inverse of the metric and we have called $\theta^{i}\equiv\theta^{R(i)}_k\dot{\epsilon}^k$
(from now on, when the script R, L is omitted we understand R). We would then write:
\[\Theta_{PC}=m\,\theta_i\theta^{(i)}-\frac{1}{2}m\,\theta_i\theta^idt\,. \]

The solutions of the equations of motion are

%(the inverse of the Hamilton-Jacobi
%transformation \cite{Conferencia-Granada})

%
\bea
\epsilon^{i}&=&\varepsilon^{i}\cos\omega t+\dot{\varepsilon}^{i}\frac{\sin\omega t}{\omega}\nn\\
\dot{\epsilon}^{i}&=&\dot{\varepsilon}^{i}\cos\omega t-\omega\varepsilon^{i}\sin\omega t \,,\label{equations}
\eea
where $\omega=\sqrt{\frac{8}{m
R^2}H}=\frac{1}{R}\sqrt{g_{ij}(\vec{\epsilon}\,)\dot{\epsilon}^{i}\dot{\epsilon}^{j}}=
\frac{1}{R}\sqrt{g_{ij}(\vec{\varepsilon}\,)\dot{\varepsilon}^{i}\dot{\varepsilon}^{j}}$;
$\vec{\varepsilon}$ and $\dot{\vec{\varepsilon}}$ denote initial
values for $\vec{\epsilon}(t)$ and $\dot{\vec{\epsilon}}(t)$. It
must be noticed that these equations are similar to those of a
harmonic oscillator but with a frequency depending on the energy.
Note that the second equation is equivalently written as
\be
\theta^i(\vec{\epsilon}(t))= \hbox{const}\equiv\vartheta^i(\vec{\varepsilon}) \;\; \hbox{(either Left- or Right-)}\label{theta}
\ee
where $\vartheta^i(\vec{\varepsilon})$ keeps the same functional dependence on $\vec{\varepsilon}$ as $\theta^i$ on $\vec{\epsilon}$.

The expression (\ref{equations}), completed with the trivial one
$t=\tau$, can be seen as an invertible transformation which goes
from variables $(t,\epsilon^i,\dot{\epsilon}^j)$ to ``constant''
coordinates and velocities
$(\tau,\varepsilon^i,\dot{\varepsilon}^j)$, in much the same way,
in the case of the free particle, the expressions $t=\tau\,,\;\;
x^i=x_0^i+\dot{x}^i_0t\,,\;\;\dot{x}^i=\dot{x}^i_0$ can be given
the same interpretation. We shall call the mentioned
transformation the ``Hamilton-Jacobi'' transformation, taking the
name from the canonical transformation that take a Hamiltonian
system to canonical constant coordinates and momenta (and null
Hamiltonian) \cite{Largo}. We refer the reader to Figure 1 in the
Appendix to visualize the transformation and the notation used for the coordinates on each space.

Equation (\ref{theta}) points the natural constants of
motion associated to the (Killing) symmetries (\ref{simetriaR}) (or
(\ref{simetriaL})), when jet-prolonged to a point symmetry
(according to the general formula (\ref{jet})), of the Lagrangian, that is
\be
X_{(i)}=Z^j_{(i)}\frac{\!\!\partial}{\partial\epsilon^j}+
\frac{\partial Z^j_{(i)}}{\partial\epsilon^k}\dot{\epsilon}^k\frac{\!\!\partial}{\partial\dot{\epsilon}^j}
=Z^j_{(i)}\frac{\!\!\partial}{\partial\epsilon^j}+\frac{2}{R}\eta^k_{.in}\theta^n\frac{\!\!\partial}{\partial\theta^k}\,.\label{simetriaKilling}
\ee
Even more, the use of $\theta^i$ instead of $\dot{\epsilon}^i$
has the advantage that the former is an intrinsic quantity
associated with the group $SU(2)$ irrespective of the particular,
local parametrization of the group (the configuration space).

A less obvious matter, however, is to realize the constants
$\varepsilon^i$ also as Noether invariants so that we can have the
Solution Manifold (SM) parameterized by intrinsic quantities. To
achieve this task, let us perform the Hamilton-Jacobi transformation on the Poincar\'e-Cartan form
(\ref{thetaPC}), that is,
(\ref{equations}) completed with the trivial transformation
$t=\tau$. After some computations we arrive at
\be
\Theta_{PC}\approx\Lambda\equiv m\vartheta_i\vartheta^{(i)}\label{Lambda}
\ee
up to a total differential (note that $\vartheta_i$ is a function
whereas $\vartheta^{(i)}$ is a $1$-form, the canonical $1$-form).
The differential $d\Theta_{PC}$ does, actually, go to SM (that is to say, it can be written in terms of functions on SM only), defining
the symplectic structure $\Omega$, generated by the Liouville
$1$-form $\Lambda$:
\be
\Omega=d\Lambda = md\vartheta_i\wedge\vartheta^{(i)}+\frac{m}{R}\vartheta_i\eta_{\cdot
jk}^{i}\vartheta^{(j)}\wedge\vartheta^{(k)}\label{Omega}\,.
\ee
The formulas above are written in intrinsic coordinates so that those
expressions are valid for any local parametrization of the
configuration space, $S^3$, as a consequence of its being a Lie
group. Nevertheless, we may resort to a particular useful
coordinates, Darboux local coordinates, where (\ref{Lambda}) and
(\ref{Omega}) adopt their canonical form:
\bea
\Lambda&=&\pi_id\varepsilon^i\nn\\
\Omega&=&d\pi_i\wedge d\varepsilon^i\label{OmegaCanonica}
\eea
where $\pi_i$ is defined by $m\vartheta_i\equiv Z^k_{(i)}\pi_k$ or $\pi_i\equiv m\vartheta_i^{(k)}\vartheta_k$.

We are now in conditions to find out the symmetries of the classical system whose Noether invariants are the
constant of motion $\varepsilon^i$. In fact, the only thing which remains is to compute the Hamiltonian vector field on SM,
$Y^{(i)}$, associated with the Hamiltonian fuction $\varepsilon^i$ according to the symplectic form (\ref{Omega}):
\be
Y^{(i)}=\frac{\!\!\partial}{\partial\pi_i}=\frac{1}{m}Z(\vec{\varepsilon})^i_{(k)}\frac{\!\!\partial}{\partial\vartheta_k}\;,\;\;\;\;
 \Omega(Y^{(i)})=d\varepsilon^i\label{Y}
\ee
To express these vectors, here obtained on SM, as symmetries of the original Poincar\'e-Cartan form (not a symmetry of
the Lagrangian, though) we must perform
the inverse Hamilton-Jacobi transformation, that is, the inverse of (\ref{equations}), completed again with the trivial
one $\tau=t$. After some rather involved calculations, we arrive at
\bea
 Y_{(j)}&=& \frac{1}{m\omega}
\left(
 (\delta^n_j-\frac{1}{R^2}\epsilon_j\epsilon^n-\frac{{\theta}^k{\theta}^s}{m^2R^2\omega^4}Z_{(k)j}Z_{(s)}^n)
 \hbox{sin}(\omega t)
+\frac{{\theta}^l{\theta}^r t }{m^2R^2\omega} Z_{(l)}^n Z_{(r)j}(\vec{\varepsilon}\,)\right)
     \frac{\!\!\partial}{\partial\epsilon^n}\nonumber\\
&+&\frac{1}{m}\left(Z_j^{(n)}\hbox{cos}(\omega t)
     +\frac{1}{\omega}
     (
      \frac{1}{R}\eta^n_{.jm}Z_{(r)}^m\theta^r + \frac{1}{R^2}\delta_j^n \theta_k\epsilon^k
     )\hbox{sin}(\omega t)
       \right)\frac{\!\!\partial}{\partial{\theta}^n}\,.\label{simetriaY}
\eea

In spite of the nearly aggressive realization of the new, boost-like, non-point symmetry on the space of evolution of the system,
the quantum treatment in terms of the basic symmetries $X_{(i)}, Y^{(j)}$, given by (\ref{simetriaKilling}), (\ref{simetriaY}),
along with their commutators, closing a finite-dimensional group, to be called $SU(2)$-Sigma group, will be quite cosy.
We shall do that in Sec. 3.2.

%Just to end the present section, let us point out the actual accordance with the general Lagrangian scheme of the previous subsection.
%In fact, defining co-ordinates $\vec{x}\equiv\frac{R}{2}\vec{\epsilon}$ (and analogous relation for $\vec{\varepsilon}$) the
Let us note that the description above clearly corresponds to that
of a particle moving on a space of constant curvature
$k=\frac{1}{R^2}$, that is, a sphere $S^3$ with radius $R$. Such
description generalizes that of the three-dimensional free
particle. In fact, if we consider the formal limit $R \to \infty$,
the symmetry generators $X_{(i)}$ (point) and $Y_{(j)}$
(non-point) become those for usual translations and boosts,
respectively:
\begin{eqnarray*}
X_{(i)} &\xrightarrow[R\to\infty]{}&  \frac{\!\!\partial}{\partial\epsilon^i}
\\
Y_{(j)} & \xrightarrow[R\to\infty]{}& \frac{\!\!\partial}{\partial\dot \epsilon^j} + t \frac{\!\!\partial}{\partial\epsilon^j} \,.
\end{eqnarray*}
This shows again that the generalization of the boosts symmetry for this non-linear case
is a non-point symmetry of the Poincar\'e-Cartan 1-form, although not a symmetry of the
Lagrangian.

\section{Non-canonical Quantization of $SU(2)$-PNL$\sigma$M}

In this Section, we take advantage of the generalized symmetry for
the $SU(2)$ particle-non-linear sigma model, describing a free
particle moving on a sphere $S^3$, just found, to build a proper
quantum description. In order to do so, we resort to the Group
Approach to Quantization algorithm, outlined in the following
Subsection.

\subsection{Group Approach to Quantization: a brief review}

We give in this section a very brief outline of the so
called Group Approach to Quantization (GAQ) or Quantization over a
Group Manifold (see \cite{Aldaya-review-GAQ} and references therein).
The basic idea of GAQ consists in taking advantage of having two mutually
commuting copies of the Lie algebra $\mathcal{\widetilde{G}}$ of
a group $\widetilde{G}$ which is a central extension of $G$  by $U\left(1\right)$) and
constitutes the basic, strict symmetry of a given physical system, that is,
\begin{equation}
\chi^{L}\left(\widetilde{G}\right)\approx\widetilde{\mathcal{G}}\approx\chi^{R}\left(\widetilde{G}\right),
\end{equation}
in such a way that one copy, let us say $\chi^{R}\left(\widetilde{G}\right)$,
plays the role of pre-quantum operators acting (by usual derivation)
on complex (wave) functions on $\widetilde{G}$, whereas the other,
$\chi^{L}\left(\widetilde{G}\right)$, is used to reduce the representation
in a manner compatible with the action of the operators, thus providing
the true quantization.

In fact, from the group law $g^{\prime\prime}=g^{\prime}\star g$
of any group $\widetilde{G}$, we can read two different actions:
\begin{eqnarray}
g^{\prime\prime} & = & g^{\prime}\star g\equiv L_{g^{\prime}}g,\\
g^{\prime\prime} & = & g^{\prime}\star g\equiv R_{g}g^{\prime}.
\end{eqnarray}
The two actions commute and so do the generators $\widetilde{Z}_{(a)}^{R}$
and $\widetilde{Z}_{(b)}^{L}$ of the left and right actions respectively,
i.e.
\begin{equation}
\left[\tilde{Z}_{(a)}^{L},\widetilde{Z}_{(b)}^{R}\right]=0\,\,\,\forall\;\; \hbox{group parameters}\;\; a,\, b.
\end{equation}
The generators $\widetilde{Z}_{(a)}^{R}$ are right-invariant vector
fields closing a Lie algebra, $\chi^{R}\left(\widetilde{G}\right)$,
isomorphic to the tangent space to $\widetilde{G}$ at the identity,
$\widetilde{\mathcal{G}}$. The same, changing $L\longleftrightarrow R$,
applies to $\widetilde{Z}^{L}\in\chi^{L}\left(\widetilde{G}\right)$.

We consider the space of complex functions $\psi$
on the whole group $\widetilde{G}$ and restrict them to only $U\left(1\right)$-functions,
that is, those which are homogeneous of degree one on the argument
$\zeta\equiv e^{i\phi}\in U\left(1\right)$, that is,
\begin{equation}
\Xi\Psi\equiv\widetilde{Z}_{(\phi)}^{L}\Psi=i\Psi,
\end{equation}
where
%$\widetilde{Z}_{(\phi)}^{L}$
$\Xi$ is the (central) generator of the
$U\left(1\right)$ subgroup. On these functions the right-invariant
vector fields act as pre-Quantum Operators by ordinary derivation.
However, this action is not irreducible since there is a set of non-trivial
operators commuting with this representation. In fact, all the left-invariant
vector fields do commute with the right-invariant ones, i.e. the (pre-quantum)
operators. According to Schur's Lemma those operators
must be ``trivialized'' in order
to turn the right-invariant vector fields into true quantum operators (that is, to achieve an irreducible representation).

We have seen that the action of the central generator
(which is, in particular, left-invariant) is fixed to be non-zero
by the $U\left(1\right)$-function condition. Thus, not every left-invariant
vector field can be nullified in a compatible way with this condition.
That is, if
\begin{equation}
\left[\tilde{Z}_{(a)}^{L},\widetilde{Z}_{(b)}^{L}\right]=\widetilde{Z}_{(\phi)}^{L},
\end{equation}
then $\widetilde{Z}_{(a)}^{L}\Psi=0$, $\widetilde{Z}_{(b)}^{L}\Psi=0$
is not compatible with $\widetilde{Z}_{(\phi)}^{L}\Psi=i\Psi$. Of
course, this null condition on $\Psi$ can be imposed by those
generators that never produce a central term by commutation, and
they constitute the {\it Characteristic sub-algebra
$\mathcal{G}_{\Theta}$}. But also half of the rest of
left-generators can be joined to $\mathcal{G}_{\Theta}$ to
constitute a {\it Polarization} subalgebra $\mathcal{P}$.

Then, the role of a polarization is that of reducing
the representation, which now constitutes a true {\it Quantization}.
Therefore we impose that wave functions satisfy the polarization condition:
\begin{equation}
\widetilde{Z}_{(b)}^{L}\Psi=0,\,\forall\widetilde{Z}_{(b)}^{L}\in\mathcal{P}.
\end{equation}

The unitarity of the representation is guaranteed by the choice of
the invariant volume on the group
\[ \mu=\theta^{L(a)}\wedge\theta^{L(b)}\wedge...\]
where $\theta^{L(a)}$ are the canonical Left-invariant $1$-forms on the group, which obviously is invariant
under the quantum operators, rendering them anti-Hermitian (actually Hermitian when multiplied by $i$).

Particularly useful becomes the canonical $1$-form in the central
direction, the {\it quantization} form
$\theta^{L(\phi)}\equiv\Theta$, which generalizes the
Poincar\'e-Cartan form of Classical Mechanics, roughly speaking
$\Theta\approx\Theta_{PC}+d\phi$. Contrarily to the
semi-invariance of $\Theta_{PC}$, the quantization form is
strictly invariant. The Characteristic sub-algebra
$\mathcal{G}_{\Theta}$ generates the Characteristic Module of
$\Theta$, that is $\widetilde{Z}/ \;\;i_{\widetilde{Z}}d\Theta=0=i_{\widetilde{Z}}\Theta$, or generalized
equations of motion. Also, the {\it Noether invariants} find their
generalization here as $i_{\widetilde{Z}^R_{(a)}}\Theta$.

It should be noticed that the choice of a given polarization subalgebra determines the particular
``representation'' of the quantum theory, that is, ``co-ordinate representation'', ``momentum representation'' or any other. In case
the Lie algebra of the quantization group does not provide a polarization subalgebra leading to the desired one, we may resort to the
left enveloping algebra to find a higher-order polarization subalgebra of the same dimension and able to accomplish our desire.

\subsection{The case of $SU(2)$-Sigma group}
\label{secpinche}

%\subsubsection{Basic symmetries and quantization group}

The basic symmetries of a dynamical system are those that close an
algebra generalizing the Heisenberg-Weyl algebra of the canonical
quantization, in the sense that it contains an even number of
operators at least, half of which play the role of
``translations'' and the other half of ``boosts''. Obviously, the
commutator of ``translations'' and ``boosts'' should provide a
central term in analogy to the canonical quantization. It is often
not easy to determine the basic symmetries of a system and let alone
if the system is not linear, as we have realized in
Subsec. 2.2.
%Although
Part of this task can be simplified, as also tested in this
example, when working in the solution manifold (SM) of the system
\cite{Conferencia-Granada}, while the realization on the evolution
space, where the Lagrangian is living, is hard and unpractical.

%We then adopt a more elegant and at the same time
%effective procedure consisting in  performing the quantization of the
%basic functions on SM and evolve the
%basic quantum operators by successive action of the Hamilton operator according to the Magnus power series. {\color{red}No es esto lo que hemos hecho, en realidad. De hecho, hemos encontrado las soluciones. Yo quitaria la segunda mitad del parrafo}

The quantization of the basic functions on SM will be achieved
along the general guides outlined above, particularized for the
finite-dimensional Lie group which arises by exponentiating the
Poisson algebra (associated with the symplectic form obtained on
SM, (\ref{Omega})) closed by $\varepsilon^i,\vartheta_j$. It is
easy to realize the closure of the following $7$-dimensional
Poisson subalgebra on SM\footnote{The Dirac-bracket method is
often used in the treatment of constrained systems, although it
is not devoid of ambiguities. In the present case, we might depart
from the canonical Poisson bracket associated with the motion of a
particle on $\mathbb{R}^4$, with global coordinates $x^\alpha,
\,\alpha,\,\beta = 1,...,4$ subjected to the constraint $x^2=R^2$. The Dirac brackets would be
\be
\{x^\alpha,\,x^\beta\}=0,\quad\{x^\alpha,\,p^\beta\}=\delta^{\alpha\beta}-
\frac{x^\alpha x^\beta}{x^2},\quad\{p^\alpha,\,p^\beta\}=\frac{1}{x^2}(p^\alpha
x^\beta-p^\beta x^\alpha)\,.
\ee
However, only the coordinates $x^i, x^4$, with $i=1,...3$, are
Noether invariants associated with globally-defined transformations
on the solution manifold; the variables $p^\alpha$ must be
replaced by global constants like, for instance,
$L^i\equiv\frac{1}{2}\eta^i_{.jk}L^{jk}+L^{i4}$, where
$L^{\alpha\beta}$ is the standard angular momentum in four
dimensions. In fact, the Lie algebra closed by $\langle L^i,\,x^j,\,x^4\rangle$
is isomorphic to our algebra closed by
$\langle\vartheta^i,\,\varepsilon^j,\,\rho\rangle$. In addition,  ambiguities
related with the definition of the quantum Hamiltonian must be
solved by hand in the Dirac-bracket method \cite{Kleinert}.}:

% In the SM the Poincaré-Cartan form (Liouville form $\varLambda$)
% takes a particularly simple expression when used Darboux coordinates,
% \begin{equation}
% \Lambda=p_{i}dx^{i}\,\,((x^{i},p_{i})\, are\, Darboux\, coordinates\, in\, SM).
% \end{equation}
% For the particle on the sphere $S^{3}$ the Liouville form is,
%
% \begin{equation}
% \Lambda=\pi_{i}d\varepsilon^{i},
% \end{equation}
% where $\varepsilon_{i}$ are the coordinates of the $SO(3)$ group
% and $\pi_{i}$ the momentum. In a previous papers \cite{Conferencia-Granada},
% the authors, have found that the following functions
% \begin{equation}
% <\varepsilon^{i},\vartheta_{i}=X_{(i)}^{k}\pi_{k},\rho=\sqrt{1-\frac{\vec{\varepsilon\,}^{2}}{4}}>;\,\,\,(here\, X_{(i)}^{k}=X_{(i)}^{L\, k}),
% \end{equation}
%  defined over the SM close a 7-dimensional Poisson algebra,
\begin{eqnarray}
\left\{ \varepsilon^{i},\varepsilon^{j}\right\}  & = & 0\nn\\
\left\{ \varepsilon^{i},\vartheta_{j}\right\}  & = & \frac{1}{R}\eta_{\cdot jk}^{i}\varepsilon^{k}+\rho\delta_{j}^{i}\nn\\
\left\{ \vartheta_{i},\vartheta_{j}\right\}  & = & \frac{2}{R}m\,\eta_{\cdot ij}^{k}\vartheta_{k}\label{Pinche}\\
\left\{ \varepsilon^{i},\rho\right\}  & = & 0\nn\\
\left\{ \vartheta_{i},\rho\right\}  & = & \frac{1}{R^2}\varepsilon_{i}\,,\nn
\end{eqnarray}
where the new function $\rho$, required to close the basic
algebra, is
\[ \rho(\vec{\varepsilon}\,)\equiv\sqrt{1-\frac{\vec{\varepsilon\,}^{2}}{R^2}}\,.
\]
The associated Hamiltonian vector fields, now denoted $X_{\vartheta_{i}}, X_{\varepsilon^j}\;(=Y^j),X_{\rho}$, and written in
terms of the standard canonical variables, are:
\begin{equation}
X_{\vartheta_{i}}=\frac{\partial\vartheta_{i}}{\partial\pi_{s}}\frac{\partial}{\partial\varepsilon^{s}}-
\frac{\partial\vartheta_{i}}{\partial\varepsilon^{s}}\frac{\partial}{\partial\pi_{s}}=
X_{(i)}^{k}\frac{\partial}{\partial\varepsilon^{k}}-\frac{\partial X_{(i)}^{k}}{\partial\varepsilon^{s}}\pi_{k}\frac{\partial}{\partial\pi_{s}},\nn
\end{equation}
\begin{equation}
X_{\varepsilon^{i}}=\frac{\partial\varepsilon^{i}}{\partial\pi_{s}}\frac{\partial}{\partial\varepsilon^{s}}-
\frac{\partial\varepsilon^{i}}{\partial\varepsilon^{s}}\frac{\partial}{\partial\pi_{s}}=-\frac{\partial}{\partial\pi_{i}},\nn
\end{equation}
\begin{equation}
X_{\rho}=\frac{\partial\rho}{\partial\pi_{s}}\frac{\partial}{\partial\varepsilon^{s}}-
\frac{\partial\rho}{\partial\varepsilon^{s}}\frac{\partial}{\partial\pi_{s}}=-\frac{1}{R^2\rho}\varepsilon_{s}\frac{\partial}{\partial\pi_{s}}.\nn
\end{equation}
This algebra generalizes and replaces the Heisenberg-Weyl or canonical algebra
for the particle over the sphere $S^{3}$ and constitutes the basic
dynamical symmetry of the system. We insist on the non-point nature of an essential part of this symmetry,
a fact that could be responsible of its being unknown, but it is as fundamental as the ordinary Heisenberg-Weyl
symmetry for linear systems.

% As far as the algebraic structure of this $SU(2)$-sigma group is concerned, it should be noted that it is really a subgroup of
% the Euclidean $E(4)$ group.

For the sake of completeness, let us mention that the symmetry (\ref{Pinche}), though being enough to achieve quantization, is
not the full symmetry of the system. In fact, it can be added with ``ordinary'' rotations generated by the Hamiltonian
functions (Noether invariants):
\[ J_i\equiv\frac{1}{2}\eta_{ijk}\varepsilon^j\pi^k \]
along with the non-independent ones
\[ \kappa_i\equiv\rho\pi_i \]
closing now an Euclidean group $E(4)$. In fact, the combinations
$J_i +\kappa_i$ and $J_i -\kappa_i$ prove to be the Noether
invariants $\theta^R_i\equiv\vartheta_i$ and $\theta^L_i$
associated with the ``right'' and ``left'' SU(2) generators,
leaving invariant the {\it chiral} Lagrangian (\ref{Lagrangian}).
This bigger (and non-minimal in the sense that it does not
constitute the minimal generalization of the Heisenberg-Weyl
group) symmetry were pointed out in \cite{Isham-Klauder,Isham} as
a possible group to undertake the quantization of the $S^3$
particle according to the Wigner-Mackey algorithm of induced
representations of semi-direct product groups
\cite{Wigner-Mackey}. There, although the realization of this
symmetry was not explicit, one would have also been lead to
non-point transformations which do not leave the Lagrangian
invariant, in contrast with the original aim.

Since we dispose of a well-defined, finite-dimensional group of symmetry
we only have to apply step by step the algorithm GAQ just described in order to write down the quantum theory.
To this end we exponentiate the algebra above and find the following
group law for the $SU(2)$-sigma group centrally extended
by $U\left(1\right)$,
\bea
\vec{\varepsilon\,}^{\prime\prime}&=&\rho\vec{\varepsilon\,}^{\prime}+\rho^{\prime}\vec{\varepsilon}+
\frac{1}{R}\vec{\varepsilon\,}^{\prime}\wedge\vec{\varepsilon}\nn\\
\vec{\nu}^{\prime\prime}&=&\vec{\nu}^{\prime}+X^{L}(\vec{\varepsilon\,}^{\prime})\vec{\nu}+\frac{1}{R}\vec{\varepsilon\,}^{\prime}z\nn\\
z^{\prime\prime}&=&z^{\prime}+\rho^{\prime}z-\frac{1}{R}\vec{\varepsilon\,}^{\prime}\cdot\vec{\nu}\\ \label{PincheGroup}
\zeta^{\prime\prime}&=&\zeta^{\prime}\zeta
e^{-i\frac{m}{\hbar}(R(\rho^{\prime}-1)z-\vec{\varepsilon\,}^{\prime}\cdot\vec{\nu})}\,,\nn
\eea
where
$\rho\equiv\rho(\vec{\varepsilon}\,),\;\rho'\equiv\rho(\vec{\varepsilon}\,')\,,\;\;
%\sqrt{1-\frac{1}{4}\vec{\varepsilon}\cdot\vec{\varepsilon}}$,
 \zeta=e^{i\phi}\in U(1)$ and the mass $m\in \mathbb{R}$ parameterizes the central extension.
 The parameters $\vec\nu$ and $z$ have the dimensions of a velocity for later convenience. A constant $\hbar$ with the
 dimensions of an action has to be introduced to keep the exponent dimensionless, although we choose units in which $\hbar=1$
 for simplicity from now on.
 The left-invariant vector fields of
the group are (we shall omit hereafter the symbol
$\;\widetilde{}\;$ over the group generators):
\bea
Z_{(\varepsilon^i)}^{L}&=&Z_{(i)}^{L\, k}\frac{\partial}{\partial\varepsilon^{k}}\nn\\
Z_{(\nu^i)}^{L}&=&Z_{(i)}^{L\, k}\frac{\partial}{\partial\nu^{k}}-
\frac{1}{R}\varepsilon_{i}\left(\frac{\partial}{\partial z}-mR\,\Xi\right)\nn\\
Z_{(z)}^{L}&=&\rho\frac{\partial}{\partial z}+\frac{1}{R}\varepsilon^{i}\frac{\partial}{\partial\nu^{i}}-m R(\rho-1)\Xi\\
Z_{(\zeta)}^{L}&=&\Xi=i(\zeta\frac{\partial}{\partial\zeta}-\bar{\zeta}\frac{\partial}{\partial\bar{\zeta}})\,,\nn
\eea
where $\Xi$ is the central generator, and the right-invariant ones:
\bea Z_{(\varepsilon^i)}^{R}&=&Z_{(i)}^{R\,
k}\frac{\partial}{\partial\varepsilon^{k}}+ \frac{1}{R}\eta_{\cdot
ik}^{j}\nu^{k}\frac{\partial}{\partial\nu^{j}}+\frac{1}{R}z\frac{\partial}{\partial\nu^{i}}-
\frac{1}{R}\nu_{i}\left(\frac{\partial}{\partial z}-m R\,\Xi\right)\nn\\
Z_{(\nu^i)}^{R}&=&\frac{\partial}{\partial\nu^{i}}\nn\\
Z_{(z)}^{R}&=&\frac{\partial}{\partial z}\\
Z_{(\zeta)}^{R}&=&\Xi=i(\zeta\frac{\partial}{\partial\zeta}-\bar{\zeta}\frac{\partial}{\partial\bar{\zeta}}).\nn
\eea
The Lie algebra (non-null) commutators for the right-invariant
vector fields are:
\begin{align}
[Z_{(\varepsilon^i)}^R,\, Z_{(\varepsilon^j)}^R]&=
-\frac{2}{R}\eta_{\cdot ij}^{k} Z_{(\varepsilon^k)}^R \nn
\\
[Z_{(\varepsilon^i)}^R,\, Z_{(\nu^j)}^R] &=
-\frac{1}{R}\eta_{\cdot ij}^{k}\,
Z_{(\nu^k)}^R+\frac{1}{R}\delta_{ij}\left(Z_{(z)}^R-m
R\,\Xi\right) \label{qpinche}
\\
[Z_{(\varepsilon^i)}^R,\, Z_{(z)}^R]
&=
-\frac{1}{R}Z_{(\nu^i)}^{R},\nn
\end{align}
to be compared with (\ref{Pinche}). The quantization 1-form $\Theta$, dual to $\Xi$, is
\be
\Theta=-m\varepsilon_{i}d\nu^{i}-m R(\rho-1)dz+\frac{d\zeta}{i\zeta},
\ee
from which we can compute the characteristic sub-algebra
\be
{\cal G}_{\Theta}=\langle\mathit{Z_{(z)}^{L}}\rangle,
\ee
indicating that the variable $z$ is not symplectic,
%or dynamics
 parametrizing a subgroup which plays a role similar to that of time in an algebraic sense.
%(which is absent from the group law (\ref{PincheGroup})).

Computing the Noether invariants we obtain:
\bea
i_{Z_{(\varepsilon^i)}^R}\Theta&=& m \left(Z_{(i)}^{Rk}\nu_k-\frac{1}{R}z\varepsilon_i\right)\nn\\
i_{Z_{(\nu_j)}^R}\Theta&=&-m\varepsilon^j\nn\\
i_{Z_{(z)}^R}\Theta&=&-m R(\rho-1)\nn
\eea

By taking explicitly quotient by $Z_{(z)}^{L}$ along with the central generator
on the group manifold, we arrive at a symplectic manifold symplectomorphic to the Solution Manifold of the Sec. \ref{particleS3}
parametrized by the Noether invariants $m\varepsilon^i$ and $m(Z_{(j)}^{Rk}\nu_k-\frac{1}{R}z\varepsilon_j) \equiv m \vartheta_j$,
%
%The central-extension parameter of the
%$SU(2)$-sigma group, our quantization group $\tilde{G}$, turns out to be $\frac{R}{2}$, and
%
where the notation $\equiv$ indicates the identification of
quantities of the quotient taken in the group and quantities in
the Lagrangian solution manifold. That quotient manifold in the
group, $\tilde{G}/[G_\Theta\otimes U(1)]$, for given $m$ and $R$,
is a symplectic manifold equivalent (symplectomorphic) to the
solution manifold of a classical particle of mass $m$ moving on
the sphere $S^3$ of radius $R$. We refer the reader to
Figure 1 in the Appendix to visualize the corresponding
quotient and the notation used for the coordinates on each space.

In order to get an irreducible representation it is necessary to impose
the polarization conditions, which are
\begin{equation}
\mathfrak{\mathcal{P}}=\langle Z_{(\nu^i)}^{L},\, Z_{(z)}^{L}\rangle.
\end{equation}
After that we arrive at wave functions of the form:
\begin{equation}
\varPsi(\zeta,\vec{\varepsilon},\vec{\nu},z)=\zeta
e^{-im(\vec{\varepsilon}\cdot\vec{\nu}+R(\rho-1)z)}\phi(\vec{\varepsilon}),
\end{equation}
where $\phi(\vec{\varepsilon})$ is an arbitrary function, save for
normalization (see Sec. 3.2.1 later). The quantum operators will
be now the right-invariant vector fields. The action on wave
functions is
\bea
i m \hat{\nu}_i\varPsi &\equiv& Z_{(\varepsilon^i)}^{R}\varPsi=\zeta
e^{-i m (\vec{\varepsilon}\cdot\vec{\nu}+R(\rho-1)z)}Z_{(i)}^{R\, k}\frac{\partial\phi(\vec{\varepsilon})}{\partial\varepsilon^{k}}\nn\\
-i m \hat{\varepsilon}_i\varPsi &\equiv& Z_{(\nu^i)}^{R}\varPsi=-i m\varepsilon_{i}\varPsi\\
-i m R \hat{\rho}\varPsi &\equiv& Z_{(z)}^{R}\varPsi=-i m R(\rho-1)\varPsi\,,\nn \eea
while their action on the wave functions restricted to
$\phi(\vec{\varepsilon})$ is
\bea
\hat{\nu}_i\phi(\vec{\varepsilon})&=&-\frac{i}{m}Z_{(i)}^{R\, k}\frac{\partial\phi(\vec{\varepsilon})}{\partial\varepsilon^{k}}\nn\\
\hat{\varepsilon}_i\phi(\vec{\varepsilon})&=&\varepsilon_{i}\phi(\vec{\varepsilon}) \label{PincheRed}\\
\hat{\rho}\phi(\vec{\varepsilon})&=&(\rho-1)\phi(\vec{\varepsilon}).\nn
\eea

%\subsubsection{Hilbert space}

% At this stage the problem is even possible to further
% reduce the representation imposing more bias condition established
% by the form of the function

Specially relevant becomes the expression of the Hamiltonian.
Given that, classically, it is written  as $\frac{1}{2m}
\delta_{ij} (m \vartheta^i) (m \vartheta^j)$ in terms of the
Noether invariants $m \vartheta^i$ (see (\ref{Hamiltoniana})), the
quantum expression is given by replacing classical invariants with
the corresponding operators, namely $\hat \nu_i$. Therefore, the
quantum Hamiltonian is unambiguously defined as $\hat{H}\equiv
\frac{1}{2}m\delta^{ij}\hat{\nu}_i\hat{\nu}_j$, which turns out to
be the Casimir operator of the $SU(2)$ subgroup:
%$\;\;$ ($k^{-1ij}=-\frac{\beta^2}{2}\delta^{ij}$)

%
\bea
\hat{H}\phi(\vec{\varepsilon})&=&-\frac{1}{2m}(Z_{(i)}^{R\, k}\frac{\partial Z_{(j)}^{R\, m}}{\partial\varepsilon^{k}}
\frac{\partial\phi(\vec{\varepsilon})}{\partial\varepsilon^{m}}+Z_{(i)}^{R\, k}Z_{(j)}^{R\, m}
\frac{\partial^{2}\phi(\vec{\varepsilon})}{\partial\varepsilon^{k}\partial\varepsilon^{m}})\nn\\
&=&-\frac{1}{2m}(-\frac{3}{R^2}\varepsilon^{m}\frac{\partial\phi(\vec{\varepsilon})}{\partial\varepsilon^{m}}+(\delta^{km}-
\frac{\varepsilon^{k}\varepsilon^{m}}{R^2})\frac{\partial^{2}\phi(\vec{\varepsilon})}{\partial\varepsilon^{k}\partial\varepsilon^{m}})\nn\\
&\equiv&-\frac{1}{2m}\Delta_{L-B}\phi(\vec{\varepsilon})=E\phi(\vec{\varepsilon})
\eea
where $\Delta_{L-B}$ stands for the Laplace-Beltrami operator associated with the metric (\ref{metrica}).

For the sake of completeness, we write the expression of the extra
operators (corresponding to the invariants $J_i-\kappa_i$, see
above) closing the Euclidean group $E(4)$ with (\ref{PincheRed}):
\be
ZL_{(\varepsilon^i)}\phi(\vec{\varepsilon})=Z_{(i)}^{L\, k}\frac{\partial\phi(\vec{\varepsilon})}{\partial\varepsilon^{k}}\label{ZLC}
\ee
They also close the $SO(4)$ subgroup of $E(4)$ with
$ZR_{(\varepsilon^i)}\equiv Z_{(\varepsilon^i)}^R$ associated with
$J_i+\kappa_i$. As a consequence, the combination

\begin{equation}
    \label{rotaciones}
    \hat{J}_i \equiv \frac{R}{2} (ZR_{(\varepsilon^i)} - ZL_{(\varepsilon^i)}) = \eta_i^{\cdot jk}\varepsilon_j \frac{\partial\;}{\partial\varepsilon^{k}}
\end{equation}
is in the Lie algebra of the group $E(4)$ and is obviously interpreted as the generators of usual rotations on the $S^3$ space.

At this point, it is interesting to consider once more the formal
limit $R\rightarrow\infty$ in the quantum operators obtained. With
the expression of $Z_{(i)}^{R\, k}$ and $Z_{(i)}^{L\, k}$ in mind
(see (\ref{simetriaR}) and (\ref{simetriaL})), it is
straightforward to check that $\hat{\nu}_i$ goes to the usual
momentum operator (over $m$), $\hat{\varepsilon}_i$ goes to the
usual position operator ($\vec\varepsilon$ is not bounded in the
limit) and the Hamiltonian $\hat H$ goes to the usual
three-dimensional free particle Hamiltonian operator, provided the
appropriate identification of coordinates is made. Also, rotations
(\ref{rotaciones}) are preserved in the limit.

From an algebraic point of view, this should not be surprising: it can be shown that there
exists an In\"on\"u-Wigner contraction \cite{Inonu} of Lie algebra (\ref{qpinche}),
corresponding to the limit $R\rightarrow\infty$, leading to the Heisenberg-Weyl algebra of the
canonical quantization $[\hat{\vec{x}},\hat{\vec{p}}\,]= i$ added with the operator $\hat{x}^2$
and rotations, i.e., a subalgebra of the Schr\"odinger algebra of symmetries of the quantum free
particle which does not include the time evolution symmetry nor dilations. Only after that
contraction, $\hat H$ closes a Lie algebra with basic operators, namely, the Schr\"odinger
 algebra.

From the previous discussion, it becomes patent again that the
construction above of the quantum theory for the free particle on
$S^3$ is a natural generalization of that for the free Galilean
particle. Let us stress that having found the relevant non-point
symmetries of the classical system is essential for such
generalization.

\subsubsection{Integration measure: Hilbert space }

In order to achieve a complete quantum description
of the system, it is necessary to define a measure on the Hilbert space.
%In the canonical approach to quantum mechanics there is no ambiguity
%free standard procedure to define the measure and is usually defined
%in a somewhat arbitrary or }\textit{\textcolor{black}{ad hoc}}\textcolor{black}{{}
%manner. The only constraint imposed on the measure is that it must
%make unitary the theory. The situation is further complicated, if
%possible, when the system is being described is non linear and the
%topology of the configuration space is non trivial. However
For functions on a Lie
group, there is a canonical integration measure: the Haar measure $\omega$. The
Haar measure for the $SU(2)$-Sigma group is given by

\begin{equation}
\omega=\theta^{L\,(\varepsilon_{1})}\wedge\theta^{L\,(\varepsilon_{2})}\wedge\theta^{L\,(\varepsilon_{3})}\wedge\theta^{L\,(\theta_{1})}\wedge\theta^{L\,(\theta_{2})}\wedge\theta^{L\,(\theta_{3})}\wedge\theta^{L\,(z)}.
\end{equation}
On polarized wave functions, in the context of GAQ, it is possible to define an invariant measure
\cite{Quasi-invariant-measure} as follows
\bea
d\mu&=&i_{X_{\theta^{1}}^{L}}i_{X_{\theta^{2}}^{L}}i_{X_{\theta^{3}}^{L}}i_{X_{z}^{L}}\omega=
\theta^{L\,(\varepsilon_{1})}\wedge\theta^{L\,(\varepsilon_{2})}\wedge\theta^{L\,(\varepsilon_{3})}=
\frac{1}{\sqrt{1-\frac{1}{R^2}\vec{\varepsilon}\cdot\vec{\varepsilon}}}\, d\varepsilon^{1}\wedge d\varepsilon^{2}\wedge d\varepsilon^{3}\\
&\equiv&\sqrt{|g|}\, d\varepsilon^{1}\wedge d\varepsilon^{2}\wedge d\varepsilon^{3}.
\eea
This measure is the Haar measure on the $SU(2)$ group as expected,
as well as the standard measure on a Riemannian  manifold ($S^3$)
with metric $g$ and determinant $|g|$. The volume on the whole
group, as is known, is
\begin{equation}
\int_{S^{3}}d\mu=2\pi^2 R^3.
\end{equation}
Finally, the scalar product between two wave functions restricted to
configuration space has the following expression
\begin{equation}
\langle\varPsi^{\prime}\mid\varPsi\rangle=\int_{S^{3}}\overline{\phi^{\prime}(\vec{\varepsilon})}\,\phi(\vec{\varepsilon})\, d\mu.
\end{equation}

All the quantum operators found in Sec. \ref{secpinche} turn out
to be self-adjoint with respect to this scalar product and,
therefore, the representation is unitary.

\subsubsection{Solutions}

We now arrive at the search for a definite basis for the (Hilbert)
space of wave functions, carrying a unitary and irreducible
representation of our symmetry group characterizing the quantum
dynamics of a particle moving on $S^3$. As usual, we choose
eigenfunctions of $\hat{H}$, but two more operators must be
simultaneously diagonalized in order to resolve the degeneracy. A
possibility (not unique) is the choice $\langle
\hat{H},\,\hat{J}^2,\,\hat{J}_3\rangle$, with
$\hat{\vec{J}}=\frac{R}{2}(ZR_{(\vec{\varepsilon})}-
ZL_{(\vec{\varepsilon})})$ as before,
%$<\hat{H},\,ZRZR_{(\varepsilon^3)},\,ZL_{(\varepsilon^3)}>$
as a maximal set of commuting observables.

To achieve the actual construction of the wave functions basis, it is convenient to resort to a hyperspherical coordinate system in the form:
\bea
\varepsilon_1&=&R\,\sin{\chi}\,\sin{\theta}\,\cos{\phi}\nn\\
\varepsilon_2&=&R\,\sin{\chi}\,\sin{\theta}\,\sin{\phi}\\
\varepsilon_3&=&R\,\sin{\chi}\,\cos{\theta}\nn\\
(\rho&=&\cos{\chi})\,,\nn
\eea
where the $\chi$ variable completes the ordinary spherical coordinates ($\theta,\, \phi$). In those variables, the required eigen-problem can be solved
with the result:
\be
\psi_{nlm}(\chi,\theta,\phi)=N_{nlm}\sin^l{\chi}\,C^{(l+1)}_{n-l}(\cos{\chi})Y_{lm}(\theta,\phi), \label{wavefunctions}
\ee
where $C^{(l+1)}_{n-l}(x)$ are the Gegenbauer polynomials in the $x$ variable, $Y_{lm}(\theta,\phi)$ are the ordinary spherical harmonics, and
$N_{nlm}$ are the following normalizing constants:
\be
N_{nlm}=2^ll!\sqrt{\frac{2(n+1)(n-l)!}{\nu(n+l+1)!}}\,.\label{normalization}
\ee
The wave functions solve the eigen-problem according to the expressions:
\bea
\hat{H}\psi_{nlm}&=&\frac{n(n+2)}{2 m R^2}\psi_{nlm}\nn\\
\hat{J}^2\psi_{nlm}&=&l(l+1)\psi_{nlm}\nn\\
\hat{J}_3\psi_{nlm}&=&m\,\psi_{nlm}\,.\nn
\eea

\bigskip

\subsubsection{``Momentum-space'' quantization}
\medskip

So far we have realized the quantization of the $S^3$-sigma particle in say the ``configuration space'' or
``coordinate representation'', since the variables upon which the wave functions depend arbitrarily are the
parameters $\varepsilon$. The ``momentum space'' can also be achieved from our quantization group by looking for
a polarization subalgebra containing the generator $X_{\varepsilon^{i}}^{L}$. Unfortunately, no first-order polarization
subalgebra does exist containing this generator and we have to seek in the left-enveloping algebra. In fact, we can construct
the following higher-order polarization subalgebra:
\be
\mathcal{P}^{HO}=\langle Z_{(\varepsilon^i)}^{L},Z_{(z)}^{LHO}\rangle
\ee
where the higher-order left generator replacing $Z_{(z)}^L$ is
\be
Z_{(z)}^{LHO}\equiv \left(Z_{(z)}^{L}\right)^2 -2i m R\,
Z_{(z)}^L+Z_{(\vec{\nu})}^L\cdot Z_{(\vec{\nu})}^L\,,
\ee
and commutes with $Z_{(\varepsilon^i)}^{L}$.
The polarization conditions on the $U(1)$-wave functions $\zeta\psi(z,\vec{\varepsilon},\vec{\nu})$ now start by
imposing the first-order condition:
\be
Z_{(\vec{\varepsilon})}^{L}\psi(z,\vec{\varepsilon},\vec{\nu})=0\;\;\Rightarrow \psi\neq\psi(\vec{\varepsilon})\,.
\ee
The condition of second order acquires a simple form after
realizing that $Z_{(z)}^{LHO}$ does commute with the entire
algebra and thus can be rewritten as $Z_{(z)}^{RHO}$, that is,
using the same algebraic expression though in terms of
right-invariant generators. We obtain:
\be \left[-2imR \frac{\!\!\partial}{\partial
z}+\frac{\partial^2}{\partial
z^2}+\frac{\partial^2}{\partial\vec{\nu}^2}\right]\psi=0\,.
\ee
With a simple redefinition of the wave function,
\be \psi(z,\vec{\nu})=e^{im R z}\phi(z,\vec{\nu}),
\label{redefinicion} \ee
intended to substract the additive constant $i$ accompanying
$\frac{\!\!\partial}{\partial z}$, we arrive at
\be
\left[\frac{\partial^2}{\partial
z^2}+\frac{\partial^2}{\partial\vec{\nu}^2}+m^2 R^2\right]\phi(z,\vec{\nu}
)=0
\ee
%
%or
%
%\[ \frac{\partial\phi}{\partialz}=\pm i\sqrt{m^2R^2+\frac{\partial^2}{\partial\vec{\nu}^2}}\,\,\phi\]
%
the solutions of which are eigen-functions of the ``Laplacian''
operator $\Delta=\frac{\partial^2}{\partial z^2}+
\frac{\partial^2}{\partial\vec{\nu}^2}$.

The solution of the polarization equations and the reduction of the
representation to the space of functions depending only on $\vec{\nu}$
(momentum) deserves further study and will be presented elsewhere.

\section{Outlook and final remarks}

%\textcolor{red}{Hablaré de:}
%\begin{enumerate}
%\item \textcolor{red}{Simetrias de contacto como algo fundamental}
%\item \textcolor{red}{Obtencion de una medida de modo natural}
%\item \textcolor{red}{Evolucion temporal del modelo}
%\item \textcolor{red}{Interacciones debiles. En teoria de campos no es necesario
%tomar traza parcial. Modelo Stueckelberg sector bosonico masivo.}
%\item \textcolor{red}{Hablar sobre el operador posicion de de Sitter.}
%\end{enumerate}

In this article we have realized a consistent quantization of a particle moving
on the sphere $S^3$ considered as the parameter space of the group $SU(2)$. It constitutes
a paradigmatic yet relatively simple non-linear (quantum-Mechanical) sigma-model
problem. To realize the proper quantization we have resorted to a
(non-canonical) Group Approach to Quantization, entirely based on symmetry
grounds and generalizing the Kostant-Kirillov-Souriau technique for Lie
group representation. The minimal, basic symmetry to achieve this task
turned out to be constituted, in part, by symmetries of the
Poincar\'e-Cartan form that do not preserve the Lagrangian, that is,
non-point symmetries; this sort of symmetries are rather well known though
mildly used in quantum theory. The use of Lie group techniques in the
quantization process of non-linear classical systems solves the order
ambiguities which inescapably arise in the Canonical Quantization scheme,
provides the adequate integration measure on non-trivial configuration spaces and a
generalization of the momentum-space ``representation'', which requires further analysis and will be
studied elsewhere. The ordinary
Canonical Quantization process naturally emerges also as group
quantization in the In\"on\"u-Wigner contraction (radius $\rightarrow\infty$)
that turns, roughly speaking, the $SU(2)$-sigma group into the Heisenberg-Weyl one.

Even though the classical system is exactly solved, we have proceeded in a more practical
way consisting in quantizing the solution manifold (SM) as associated with the
$SU(2)$-sigma group (as a co-adjoint orbit) and introducing then the Hamiltonian operator
as an unambiguous function of the already represented basic generators of the symmetry group.

We should mention that the present quantized system can also serve
as a toy model for future proper quantization of $SU(2)$-sigma
models in field theory. The ``local'' $SU(2)$-sigma group was
already presented, at the Lie algebra level, in a previous paper
\cite{Conferencia-Granada} and it would be intended to provide the
adequate treatment of the bosonic sector in Non-Abelian
Stueckelberg (massive-gauge) theories of interactions.

%shown the inescapable need to consider non-point
%symmetry transformations when describing both classical and quantum
%non-linear systems. We are convinced that its this lack of open-mindedness
%that prevents a considerable advance in the quantum descriptions of
%nonlinear problems.

%Often it is considered, especially in field theory, point symmetries
%(geometric symmetries) as the only relevant. Not so in particle mechanics
%\cite{Goldstein} where the use of contact transformations
%is mildly used. From the mathematical point of view the theory of
%particles and the theory of fields use the same tools and share the
%same structures, therefore, logical thoughts about using the same
%techniques in one case and in another. Although we have been trying
%elementary systems, i.e., non macroscopic, the techniques of description
%using symmetries are not at all restricted to that field. Other branches
%of physics like fluid mechanics

\section*{Acknowledgments}

Work partially supported by the  Spanish MINECO, Junta de
Andaluc\'\i a and Fundaci\'on S\'eneca,  under projects
FIS2014-57387-C3-1P and FIS2014-57387-C3-3P, FQM219-FQM4496 and
08814/PI/08, respectively.

\section*{Appendix}
\label{mapa}

We provide here a diagram in which the notation for coordinates in each space is specified. Note that the diagram refers
to the classical description, although the notation for the group is maintained for the quantum description.

\begin{figure}[H]

\begin{tikzpicture}[node distance=6cm]

\node (arribaL) [arriba1] {

\textbf{Lagrangian}
$\phantom{\epsilon^i(t),\dot{\epsilon}^i(t)}$
$\epsilon^i(t),\dot{\epsilon}^i(t)$
$\phantom{.}\qquad p_i(t) \qquad$
$\phantom{.}\qquad \theta^i(t) \qquad$
$\phantom{\epsilon^i(t),\dot{\epsilon}^i(t)}$
};

\node (arribaG) [arriba2, right of=arribaL, xshift=4cm] {

\textbf{Group}
$\phantom{\epsilon^i(t),\dot{\epsilon}^i(t)\qquad.}$
$\phantom{.}\qquad\varepsilon^i\qquad$
$\phantom{.}\qquad\nu^i\qquad$
$\phantom{.}\qquad z \qquad$
$\phantom{.}\qquad \zeta \qquad$
};

\node (abajoL) [abajo1, below of=arribaL, yshift=-1cm, xshift=0.5cm] {

\textbf{Lagrangian SM}
$\phantom{\qquad \epsilon^i(t),\dot{\epsilon}^i(t)}$
$\phantom{.}\qquad\varepsilon^i,\dot{\varepsilon}^i \qquad$
$\phantom{.}\qquad \pi_i \qquad$
$\phantom{.}\qquad \vartheta^i \qquad$
$\phantom{\epsilon^i(t),\dot{\epsilon}^i(t)}$
};

\node (abajoG) [abajo2, below of=arribaG, yshift=-1cm, xshift=-0.5cm] {

\textbf{Group/z}
$\phantom{\qquad \epsilon^i(t),\dot{\epsilon}^i(t)}$
$\phantom{.}\qquad\varepsilon^i\qquad$
$\phantom{.}\qquad \pi_i \qquad$
$\phantom{.}\qquad \vartheta^i \qquad$
$\phantom{\epsilon^i(t),\dot{\epsilon}^i(t)}$
};

\draw [flecha] (arribaL) --  node[text width=3cm,anchor=east] {Hamilton-Jacobi transformation} (abajoL);

\draw [flecha] (arribaG) -- node[text width=1.5cm,anchor=west] {quotient by $z$, $\zeta$} (abajoG);

\draw [ident] (abajoL) -- node[anchor=north] {identification} (abajoG);
\label{figura}

\end{tikzpicture}

\caption{Notation for coordinates in Lagrangian mechanics and Group Manifolds}
\end{figure}
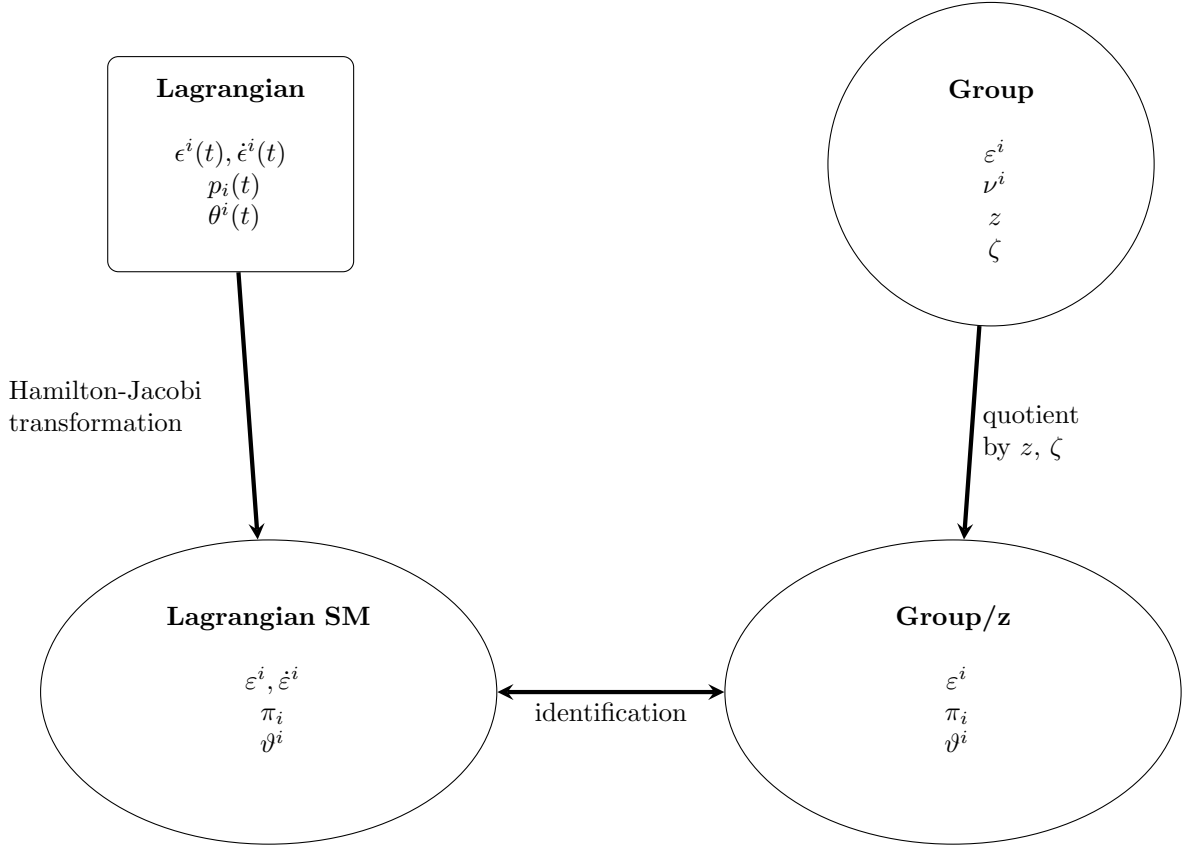

\end{document}